\documentclass[aps,prl,twocolumn,superscriptaddress,noshowpacs]{revtex4}
\usepackage{amsmath}
\usepackage{amssymb}
\usepackage{tikz}
\usepackage{graphicx}
\usepackage{dcolumn}
\usepackage{bm}
\usepackage{epsfig}
\usepackage{psfrag}
\usepackage{latexsym}

\newcommand{\refsec}[1]{\mbox{Sec.~\ref{#1}}}
\newcommand{\reftab}[1]{\mbox{Tab.~\ref{#1}}}

\newcommand{\Pb}  {\mbox{$^{208}${\rm Pb}}}
\newcommand{\bF}{\mathbf}
\newcommand{\ltapprox}  {\mbox{${ {}^<\!\!\!\!_\sim}$}}
\newcommand{\cfg}   {con\-figu\-ra\-tion}

\newcommand{\expt}  {ex\-peri\-ment}
\newcommand{\exE}   {ex\-citation energy}

\newcommand{\exEs}  {ex\-citation energies}

\newcommand{\oph}   {one-par\-ticle one-hole}
\newcommand{\tph}   {two-par\-ticle two-hole}
\newcommand{\coef}  {co\-efficient}
\begin{document}
\title{ Chaos and regularity in the doubly magic nucleus $^\bF{208}$\bf{Pb}
}
\author{B. Dietz}
\altaffiliation{
  email:  Dietz@ikp.tu-darmstadt.de}
\affiliation{%
School of Physical Science and Technology, and Key
Laboratory for Magnetism and Magnetic Materials of MOE, Lanzhou University,
Lanzhou, Gansu 730000, China
}
\author{A. Heusler}
\altaffiliation{
  email: A.Heusler@mpi-hd.mpg.de
}
\affiliation{%
  Gustav-Kirchhoff-Strasse 7/1,
  D-69120 Heidelberg, Germany
}
\author{K. H. Maier
}
\altaffiliation{deceased}
\affiliation{%
  Institute of Nuclear Physics, Polish Academy of Sciences, 
  31-342 Krak\'ow, Poland
}
\author{A. Richter}
\altaffiliation{email: Richter@ikp.tu-darmstadt.de }
\affiliation{%
  Institut f\"ur Kernphysik,
  Technische Universit\"at Darmstadt,
  D-64289 Darmstadt, Germany
}
\author{B. A. Brown}
\altaffiliation{email: brown@nscl.msu.edu}
\affiliation{Department of Physics and Astronomy and National Superconducting
Cyclotron Laboratory
Michigan State University, East Lansing,
Michigan 48824-1321, USA}
\date{\today}
\bigskip
\begin{abstract}
  High resolution \expt s have recently lead to a complete
  identification (energy, spin, and parity) of 151 nuclear levels up
  to an \exE\ of $E_x= 6.20$\,MeV in \Pb\ [Heusler {\it et al.},
  Phys.\ Rev.\ C {\bf 93}, 054321 (2016)].  We present a thorough study of the fluctuation properties in the energy spectra  
  of the unprecedented set of nuclear bound states.  In a first approach we grouped states with
  the same spin and parity into 14 subspectra, analyzed standard statistical measures for short- and long-range correlations, i.e., the nearest-neighbor spacing distribution, the number variance $\Sigma^2$, the Dyson-Mehta $\Delta_3$ statistics, and the novel distribution of the ratios of consecutive spacings of adjacent energy levels in each energy sequence and then computed their ensemble average. Their comparison with a random matrix ensemble which interpolates between Poisson statistics expected for regular systems and the Gaussian Orthogonal Ensemble (GOE) predicted for chaotic systems shows that the data are well described by the GOE. In a second approach, following an idea of Rosenzweig and Porter [Phys.\ Rev.\ {\bf 120}, 1698 (1960)] we considered the complete spectrum composed of the independent subspectra. We analyzed their fluctuation properties using the method of Bayesian inference involving a quantitative measure, called the chaoticity parameter $f$, which also interpolates between Poisson ($f=0$) and GOE statistics ($f=1$). It turns out to be $f\approx 0.9$. This is so far the closest agreement with GOE observed in spectra of bound states in a nucleus.  The same analysis has also been performed with spectra computed on the basis of shell model calculations with different interactions (SDI, KB, M3Y). While the simple SDI exhibits features typical for nuclear many-body systems with regular dynamics, the other, more realistic interactions yield chaoticity parameters $f$ close to the experimental values.
\end{abstract}
\maketitle
\begin{figure}[ht!] 
{ \includegraphics[width=\linewidth]{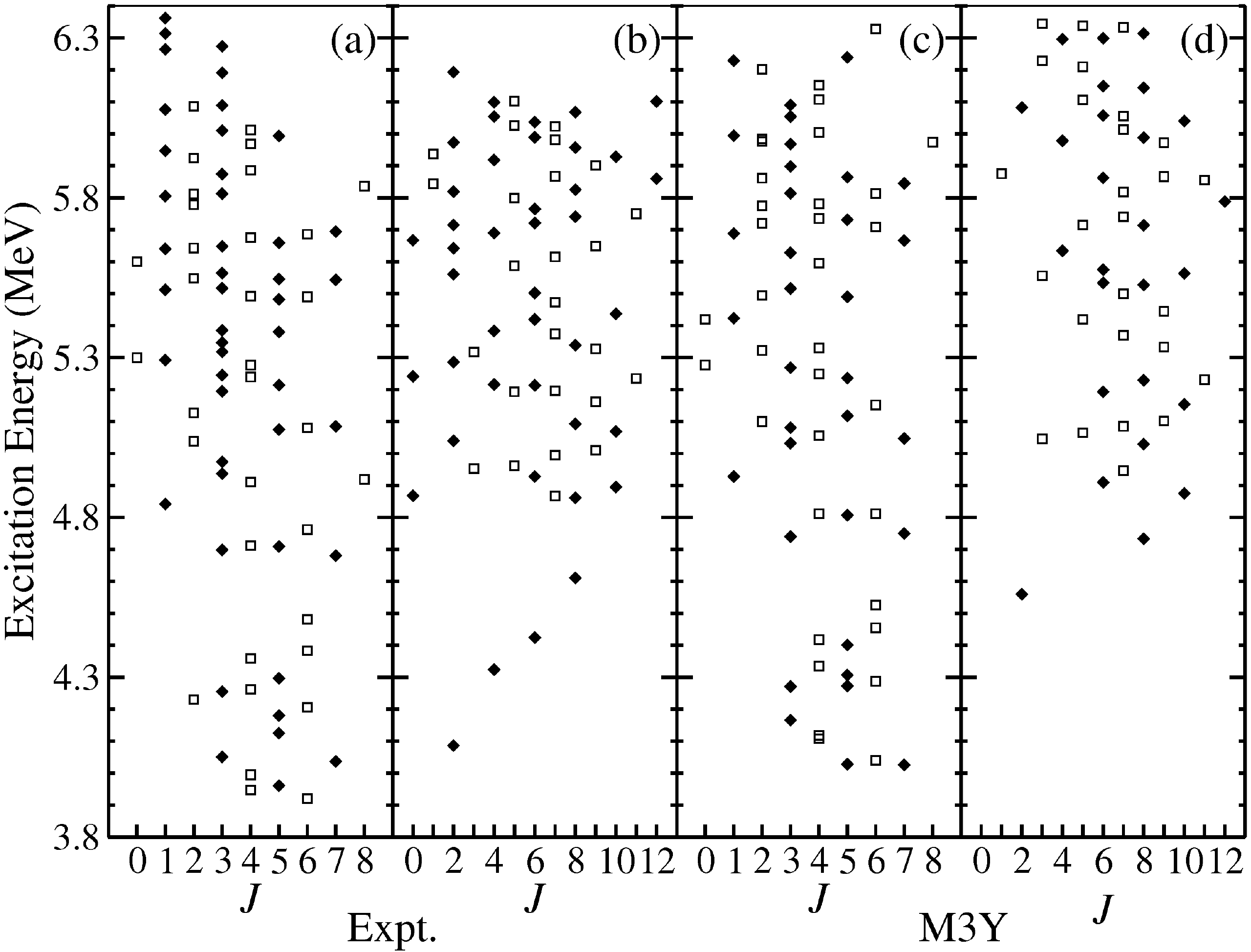}}
    \caption{ Experimental levels for states in \Pb\ at $3.8<E_x<6.4$\,MeV with negative parity (a) and positive parity (b). They are compared to the levels calculated for the M3Y residual interaction with negative parity (c) and positive parity (d). Filled diamonds and open squares show the levels with natural and unatural parity, respectively.
      \label{fig.BDzARi}
    }
  \end{figure}

  {\it Introduction}.--
  The stable doubly magic nucleus \Pb\ is one of the most studied
  nuclei both \expt ally and theoretically.  Many of its spectral
  properties are basically understood in terms of the
  nuclear shell model.  In recent years, however, a number of high
  resolution \expt s using various types of nuclear reactions have
  been performed~\cite{P.All6.2MeV, NDS2007, P.s1p3, 11th_Heu}.  The
  foremost result is that now the complete level scheme in \Pb\ is
  established up to an \exE\ of $E_x=6.20$\,MeV comprising 151 bound states, of which the  energy, spin, and parity have been unambiguously determined~\cite{P.All6.2MeV}. More states with spin and parity $J^\pi=1^-,
  2^-, 3^-$ are known up to $E_x\approx 7.5$\,MeV~\cite{NDS2007,
    P.s1p3, 11th_Heu}. In Fig.~\ref{fig.BDzARi} the \expt al levels are shown separately for 
  states of negative (a) and positive (b) parity for a range of \exEs\ $3.8<E_x< 6.40$\,MeV. 
Levels corresponding to states with natural and unnatural parity are shown as filled diamonds 
and open squares, respectively. 

The level scheme of \Pb\ has also been the subject of many shell
  model calculations of the \oph\ and \tph\ type~\cite{KuoBrown1968,
    KuoB66, Kuo1967, Kuo197093, PhysRevC.3.2421, Gillet196444,
    2000Brwn, KHMaier2007, Bertsch1977, PhysRevC.26.1323,
    P.mSM_Jolos, Mosz1979, Talmi1993}.
  Mostly, the coupling of proton particles in the orbitals
  $1h_{9/2}$,  $2f_{7/2}$,  $2f_{5/2}$,  $3p_{3/2}$,  $3p_{1/2}$,  $1i_{13/2}$ 
  to proton holes in the orbitals
  $1g_{7/2}$,  $2d_{5/2}$,  $2d_{3/2}$,  $3s_{1/2}$,  $1h_{11/2}$, 
  and the coupling of neutron particles in the orbitals
  $1i_{11/2}$,  $2g_{9/2}$,  $2g_{7/2}$,  $3d_{5/2}$,  $3d_{3/2}$,  $4s_{1/2}$,   $1j_{15/2}$ 
  to neutron holes in the orbitals
  $1h_{9/2}$,  $2f_{7/2}$,  $2f_{5/2}$,  $3p_{3/2}$,  $3p_{1/2}$, $1i_{13/2}$ 
  were taken into account; see, e.g., Fig.~1 of Ref.~\cite{Warburton1991}, however, note the different labelings used for the main quantum number. For detailed level schemes of the relevant
  proton and neutron orbits in the four neighboring nuclei of \Pb\
  see, e.g., Fig.~1 in~\cite{P.All6.2MeV} or Fig.~3-3 in~\cite{BM1969}.  The calculations of Kuo and Brown (KB) use four additional orbitals~\cite{KuoBrown1968, KuoB66, Kuo1967, Kuo197093}.
  In total, there are 27 different combinations of spin and parity for \oph\ states in \Pb. The surface-delta interaction (SDI), which acts only at the nuclear surface, has been introduced as a simple extension of the schematic shell model without residual interaction. The associated interaction strength depends on the atomic mass number as the ratio of the surface to the volume term and on geometrical recoupling \coef s~\cite{Mosz1979, Talmi1993,P.mSM_Jolos}. The KB interaction is based on the free nucleon-nucleon potential of Hamada and Johnston~\cite{KuoBrown1968, KuoB66, Kuo1967, Kuo197093} and the more recent two-body Michigan-three-Yukawas (M3Y) interaction on a one-boson exchange potential with short-range components determined with the help of the Reid nucleon-nucleon potential~\cite{KHMaier2007, Bertsch1977, PhysRevC.26.1323}. 
Figure~\ref{fig.BDzARi} shows the excitation energies calculated with the shell model which employs the effective nucleon-nucleon interaction M3Y for negative (c) and positive (d) parity. 

The main objective of the present Letter is {\it not} a level by level comparison, but rather a comparison of the spectral properties of the whole set of detected bound states in \Pb\ up to an \exE\ of $E_x=6.20$\,MeV, which is still about one MeV below the neutron threshold ($S(n)=7.368$\,MeV), and of the three theoretical models with Random Matrix Theory (RMT) results. For a generic quantum system with classically regular dynamics the spectral properties are predicted to coincide with those of Poissonian random numbers~\cite{Berry1977}.  According to the Bohigas-Giannoni-Schmit conjecture~\cite{PhysRevLett.52.1} the spectral properties of time-reversal invariant chaotic systems are well described by those of the eigenvalues of real-symmetric matrices with Gaussian distributed random entries from the GOE~\cite{McDonald1979, Casati1980, Berry1981, Dyson1963, Mehta1990}. These features are also observed in nuclear many-particle systems with no obvious classical analogue. Their spectral properties are described by Poissonian statistics, if the motion of the particles is collective, whereas for sufficiently complex motion they exhibit GOE statistics~\cite{Guhr1989,Flambaum1994,Zelevinsky1996,Gomez2011}. There are various methods to obtain information on the chaoticity vs.\ regularity in a nuclear many-body system from its spectrum, see e.g.\ the review articles~\cite{RevModPhys.81.539,Gomez2011}. We analyzed the fluctuation properties of the energy levels using two models, where one is based on a RMT ensemble~\cite{Lenz1991,Alt1993,Kota2014} and the other one on the method of Bayesian inference~\cite{Abul1996, Abul2004, Abul2006}. Both provide quantitative measures for the chaoticity in terms of a parameter which interpolates between the Poisson statistics and GOE statistics.

  {\it Spectral data}.-- 
 We note that besides for $^{208}$Pb, there are to our knowledge only two other nuclei for which complete level schemes exist, namely $^{26}$Al and $^{30}$P~\cite{Mitchell1988,Shriner2005,1402-4896-2001-T90-015}.  That of $^{116}$Sn has been termed ``nearly complete'' by the authors of Ref.~\cite{Raman1991,Garrett1997}.  Furthermore, the analysis of the spectral properties of experimentally determined bound states of nuclei did not yield the results expected in nuclear many-body systems exhibiting a chaotic dynamics. For light nuclei this was attributed to the insufficient number of identified bound states, and for nuclei like $^{26}$Al and $^{30}$P to the partial isospin symmetry breaking~\cite{Mitchell1988,Shriner2005}. The present analysis of the \expt al data is based on the 151 bound states with unambiguously assigned parity and spin~\cite{P.All6.2MeV}, so the results are statistically significant. 
  
We applied two different approaches to analyze the spectral fluctuation properties of the \expt al data~\cite{P.All6.2MeV, NDS2007, P.s1p3, 11th_Heu} and calculations~\cite{KuoBrown1968, KuoB66, Kuo1967, Kuo197093, PhysRevC.3.2421, Gillet196444, 2000Brwn, KHMaier2007, P.mSM_Jolos, Mosz1979, Talmi1993}. In the first approach we determined the ensemble averages of the statistical measures for the spectral properties of each level sequences~\cite{Bohigas1983} characterized by spin and parity.  In the second approach we followed an idea of Rosenzweig and Porter~\cite{PhysRev.120.1698} and considered the complete spectrum composed of the independent subspectra and analyzed their fluctuation properties using the method of Bayesian inference. Note, that the spectral properties are Poissonian, when we use the complete experimental spectrum irrespective of parity and spin, see Fig.~6 of~\refsec{Supplemental}. In both approaches we considered only subspectra that contained at least 5 levels. Furthermore, we first unfolded the energy levels $E_i$ in each sequence individually by replacing them by the smooth part $\bar N(E_i)$ of the integrated level density, yielding a mean spacing of unity, $\langle s\rangle =1$. The latter was determined from a fit of a third order polynomial to the integrated level density. In order to assure ourselves, that the spectral properties do not depend on the unfolding procedure, we, in addition unfolded the energy levels using an empirical formula~\cite{Cameron1965,Gilbert1965}, $\bar N (E)=\exp((E-E_0)/T)+N_0$ with $T,\ E_0$ and $N_0$ the fit parameters. It was applied to low-lying nuclear levels in Ref.~\cite{Shriner1991}. We came to the result that both procedures yield very similar results for the fluctuations of the energy levels. 

  {\it Ensembles of subspectra}.--
In order to get information on the chaoticity of the nuclear many-body system we
  calculated for each sequence of unfolded levels
  (i) the nearest-neighbor spacing distribution (NNSD) $P(s)$,
  (ii) the number variance $\Sigma^2$, 
  (iii) the Dyson-Mehta $\Delta_3$ statistics~\cite{Dyson1963,Mehta1990} which
  gives the least-square deviation of the integrated level density 
  from the best-fit straight line, and (iv) the distribution of the ratios of the consecutive spacings of adjacent energy levels~\cite{Atas2013,Dietz2016} which has the advantage that the energy levels need not be unfolded. The corresponding analytical expressions for Poisson and GOE statistics are given in Refs.~\cite{Dyson1963,Mehta1990,Atas2013}. 

\begin{figure}[ht!]
      {\includegraphics[width=0.9\linewidth]{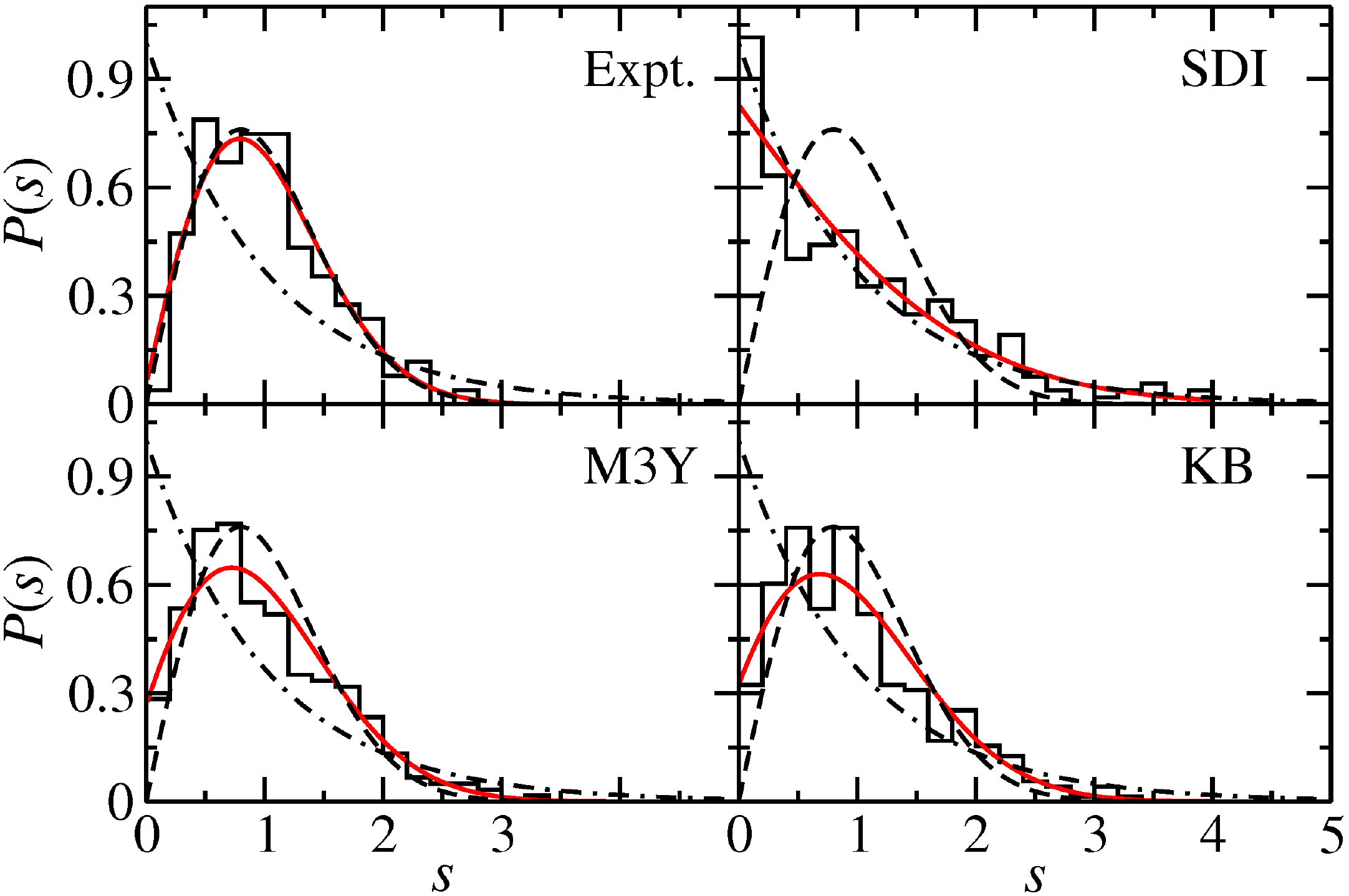}} 
    \caption{(color online) Nearest-neighbor spacing distribution
      of all experimental and calculated energy levels, respectively (histograms).  They are compared to the
      Poisson (dash-dotted line) and the GOE (dashed line) distribution.
     The full curves in red were determined with the
      method of Bayesian inference [ Eqs.~(\ref{eq.NNDComp}) and~(\ref{PsBayes})].
    \label{fig2}
  }
  \end{figure}

\begin{figure}[ht!]
      {\includegraphics[width=0.9\linewidth]{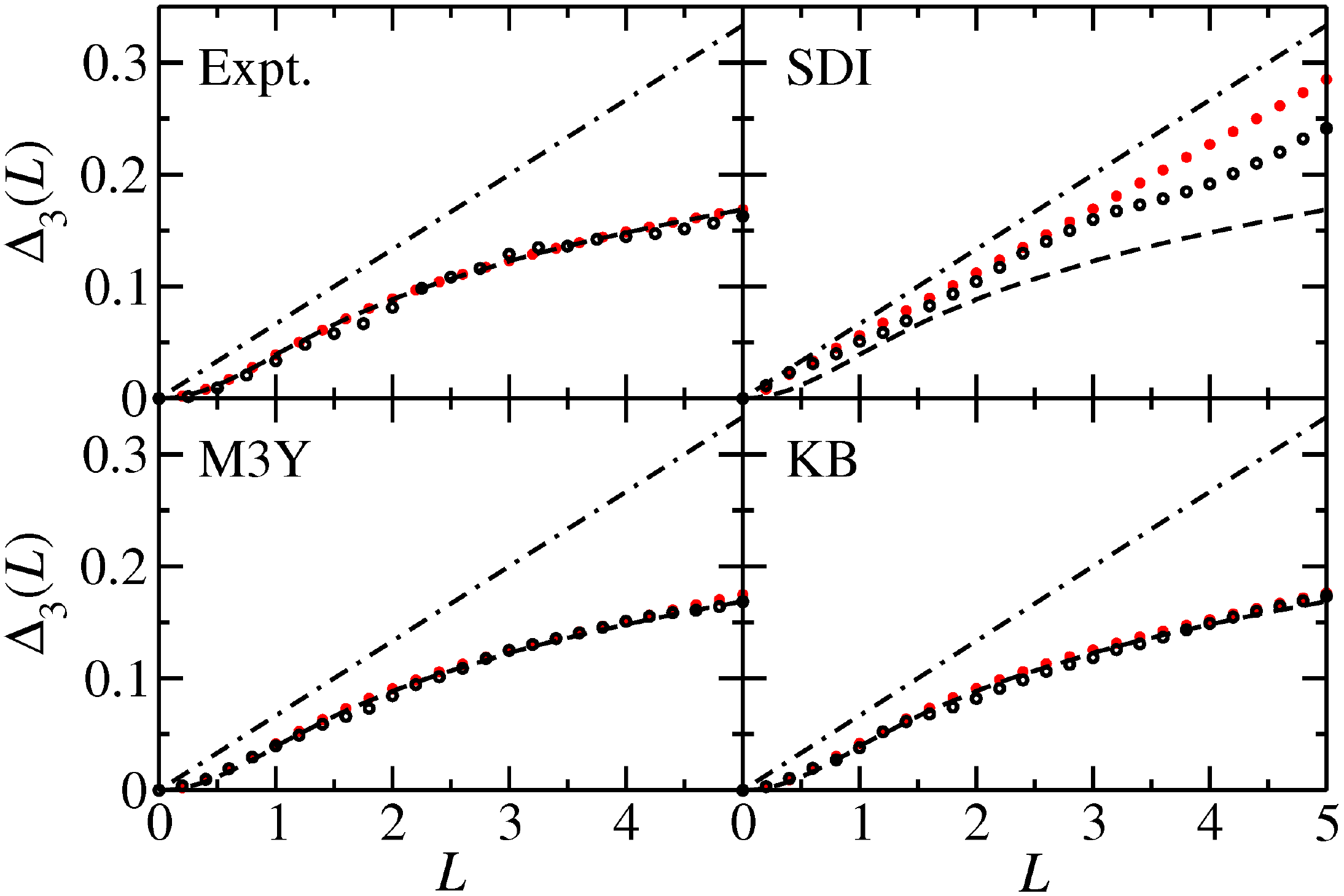}} 
    \caption{(color online) The Dyson-Mehta statistics of all experimental and calculated energy levels, respectively (black circles) in comparison to the Poisson (dash-dotted line) and GOE (dashed line) results, and the RMT model Eq.~(\ref{PGOE}) for the corresponding best-fit parameter $\lambda$ (red dots).
      \label{fig3}
    }
  \end{figure}

\begin{figure}[ht!]
      {\includegraphics[width=0.9\linewidth]{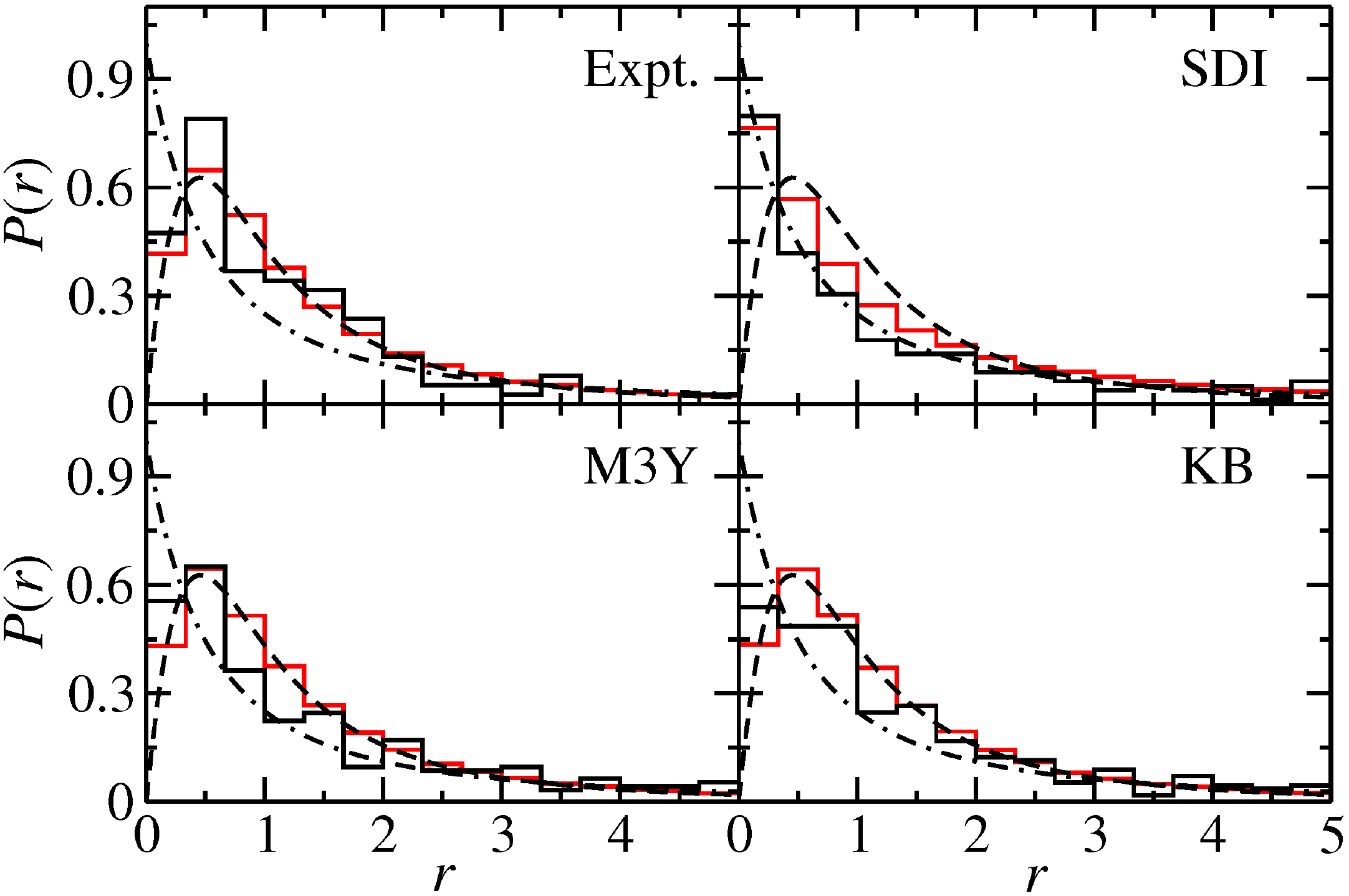}} 
    \caption{(color online) Ratio distributions of all experimental and calculated energy levels, respectively (black histogram) in comparison to the Poisson (dash-dotted line) and GOE (dashed line) results, and the RMT model Eq.~(\ref{PGOE}) for the corresponding best-fit parameter $\lambda$ (red histograms).
      \label{fig4}
    }
  \end{figure}

We analyzed the statistical measures for the \expt al energy levels and for the levels obtained from nuclear model calculations using the SDI, M3Y, and KB interactions for each of the $m$ subspectra separately and subsequently computed their ensemble averages, where the values of $m$ are given in~\reftab{tab.table1}. Then we compared them to those of random matrices from an ensemble interpolating between Poisson and GOE~\cite{Lenz1991,Kota2014},
\begin{equation}
H(\lambda)=(H_0+\lambda H_1)/\sqrt{1+\lambda^2},
\label{PGOE}
\end{equation}
where $H_0$ is a diagonal matrix of random Poissonian numbers and $H_1$ is a matrix from the GOE. Here, the range of values of the entries of $H_0$ coincided with that of the eigenvalues of $H_1$. Furthermore, the variances of the matrix elements of $H_0$ and $H_1$ were chosen such that the mean spacings of their eigenvalues equaled unity, respectively. For $\lambda=0$, the statistics is Poissonian, whereas for $\lambda\gtrsim 1$ the statistical properties are close to those of random matrices from the GOE. In Ref.~\cite{Lenz1991} a Wigner-like approximation was derived for the NNSD using $2\times 2$ random matrices. It is given in terms of the Bessel function $I_0(x)$ and the Kummer function $U(a,b,x)$ as
\begin{eqnarray}
P_{P\to GOE}(s,\lambda)&=&su(\lambda)^2/\lambda \exp\left[-u(\lambda)^2s^2/4\lambda^2\right]\nonumber\\
&\times &\int_0^\infty d\xi e^{-\xi^2-2\xi\lambda}I_0(\xi su(\lambda)/\lambda),
\label{Lenz}
\end{eqnarray}
with $u(\lambda)=\sqrt{\pi}U(-1/2,0,\lambda^2)$. In order to determine the parameters $\lambda$ for the experimental and calculated spectra, we fit this expression to their NNSDs and also compared their $\Sigma^2$-statistics with that obtained for the RMT model Eq.~(\ref{PGOE}) around the respective $\lambda$ values. Inclusion of long-range correlations turned out to be crucial for the determination of the best-fit parameters. The $\lambda$ values and the mean-square deviation of the respective NNSD from the analytical one, $\sigma_\lambda$, are given in the last two columns in the rows termed ``all'' in Table \ref{tab.table1}. We repeated the analysis taking into account only levels with positive, negative, natural ($J^\pi= 1^-, 2^+, \dots$) and unnatural ($J^\pi=0^-, 1^+, 2^-, \dots$) parity, respectively. This analysis clearly revealed that the spectral properties of the experimental levels are very close to GOE. The NNSDs for the M3Y and the KB interaction models are also close to GOE, however, their $\Sigma^2$ statistics hint at a slightly larger contribution from regular behavior than for the experimental ones. Interestingly, for all these cases the spectra of levels with positive or natural parity are closer to GOE than those with negative and unnatural parity, respectively. The SDI model, on the contrary, clearly exhibits Poissonian features. We illustrate the findings in Figs.~\ref{fig2}-\ref{fig4} where we compare the NNSD, the $\Delta_3$ statistics and the ratio distributions of the experimental and calculated levels (histograms and circles) with those of Poissonian random numbers (dash-dotted lines) and of random matrices from the GOE (dashed lines). Furthermore, Figs.~\ref{fig3} and~\ref{fig4} show the results (red histogram and dots) obtained from the RMT model Eq.~(\ref{PGOE}) using the $\lambda$ values given in Table~\ref{tab.table1}; see also Figs. 1-3 of Ref.~\refsec{Supplemental}. A second measure for the chaoticity vs. regularity is analyzed in the next section.
  \begin{table}[t!]
    \caption{The number of sets $m$ and spacings $N$ for the composite
      \expt al spectra (all), positive ($+$) and negative
      ($-$), natural (nat.) and unnatural (unnat.) parity, and for those calculated with three
      different models. The numbers $N$ are larger for the models than for the experimental data, because we had levels with energies larger than 6.2~MeV at our disposal. The chaoticity parameter and its variance $f=\overline{f} \pm\sigma$ was obtained with the method of Bayesian
      inference [Eqs.~(\ref{eq.NNDComp}) and~(\ref{PsBayes})]. The chaoticity parameter $\lambda$ was obtained 
      with the RMT model Eq.~(\ref{PGOE}) and $\sigma_\lambda$ gives the mean-square deviation of the respective NNSD from the
      corresponding analytical expression Eq.~(\ref{Lenz}).
      \label{tab.table1}
    }
 \begin{tabular}
      {@{\extracolsep{00pt}}
        @{\extracolsep{10pt} }{l}
        @{\extracolsep{10pt} }{c}
        @{\extracolsep{10pt} }{r}
        @{\extracolsep{10pt} }{r}
        @{\extracolsep{10pt} }{c}
        @{\extracolsep{10pt} }{c}
        @{\extracolsep{10pt} }{c}
        @{\extracolsep{10pt} }{c}
      }%
      \hline\hline
      \noalign{\smallskip}
      {\rm Model} & Parity & $m$ & $N$ & $\overline{f}$ & $\sigma$ & $\lambda$ & $\sigma_\lambda$\\
      \noalign{\smallskip}
      \hline
      {\rm Expt.} &     all & 14& 128& 0.95 & 0.015 & 1.50 & 0.060 \\
      {\rm SDI}   &     all & 13& 262& 0.18 & 0.069 & 0.08 & 0.001 \\
      {\rm M3Y}   &     all & 13& 282& 0.73 & 0.066 & 0.64 & 0.053 \\
      {\rm  KB}   &     all & 14& 257& 0.62 & 0.070 & 0.60 & 0.074 \\
\noalign{\smallskip} %
      {\rm Expt.} &     $+$ & 7 & 45 & 0.94 & 0.033 & 1.70 & 0.140 \\
      {\rm SDI}   &     $+$ & 6 & 82 & 0.05 & 0.042 & 0.01 & 0.010 \\
      {\rm M3Y}   &     $+$ & 6 & 84 & 0.84 & 0.079 & 1.13 & 0.081 \\
      {\rm  KB}   &     $+$ & 4 & 62 & 0.88 & 0.059 & 1.50 & 0.118 \\
\noalign{\smallskip} %
      {\rm Expt.} &     $-$ & 7 & 83 & 0.87 & 0.091 & 0.70 & 0.093 \\
      {\rm SDI}   &     $-$ & 7 & 180& 0.10 & 0.062 & 0.05 & 0.025 \\
      {\rm M3Y}   &     $-$ & 7 & 198& 0.66 & 0.085 & 0.64 & 0.057 \\
      {\rm  KB}   &     $-$ & 10& 295& 0.62 & 0.070 & 0.50 & 0.079 \\
\noalign{\smallskip} %
      {\rm Expt.} &  nat.& 5 & 79 & 0.92 & 0.051 & 1.20 & 0.096 \\
      {\rm SDI}   &  nat.& 7 & 136& 0.16 & 0.087 & 0.05 & 0.001 \\
      {\rm M3Y}   &  nat.& 7 & 147& 0.80 & 0.073 & 0.75 & 0.094 \\
      {\rm  KB}   &  nat.& 7 & 169& 0.74 & 0.091 & 1.10 & 0.085 \\
\noalign{\smallskip} %
      {\rm Expt.} &unnat.& 6 & 49 & 0.89 & 0.084 & 2.00  & 0.120 \\
      {\rm SDI}   &unnat.& 6 & 126& 0.09 & 0.063 & 0.10  & 0.001 \\ 
      {\rm M3Y}   &unnat.& 6 & 135& 0.65 & 0.096 & 0.58  & 0.089 \\
      {\rm  KB}   &unnat.& 6 & 188& 0.62 & 0.088 & 0.50  & 0.079 \\
      \hline\hline
    \end{tabular}
  \end{table}

  {\it Superimposed subspectra}.--
For the analysis of the composite spectra we proceeded as described in~\cite{Abul1996, Abul2004, Abul2006}. Accordingly, we first computed the spacings between adjacent unfolded energy levels in each subspectrum separately, and then merged them irrespective of spin and parity into one sequence of spacings $s_i,\, i=1, \dots, N$ with $N$ given in~\ref{tab.table1}. An approximate expression was derived for the NNSD $p(s,f_1, \dots, f_m)$ of a spectrum composed of $m$ subspectra with fractional level numbers $0<f_j\leq 1,\, j=1, \dots, m$ in Ref.~\cite{PhysRev.120.1698}. In Ref.~\cite{Abul1996} an approximation was derived which depends only on one parameter $f = \sum\nolimits_{j=1}^m f_j^2$ and is given by
  \begin{equation} 
    \label{eq.NNDComp}
    p(s,f) =
    \left(1 -f + 
      \frac {\pi} {2}
      Q(f) s \right)
     \exp\left(
    -(1 - f) s - 
    \frac {\pi} {4} Q(f) s^2
     \right) 
  \end{equation}
with $Q (f)=0.7 f +0.3f^2$.  Note, that for $f=0$, which corresponds to a spectrum composed of a large number of subspectra consisting of a few number of levels, this distribution approaches the Poisson distribution, whereas for $f\to 1$ it converges to the NNSD of the GOE.  Therefore, $f$ is referred to as chaoticity parameter.  

In order to determine the values of $f$ for the experimental and calculated composite spectra, we used the method of Bayesian inference~\cite{Harney2003}.  Assuming that the spacings $s_j,\, j=1, \dots, N$ are statistically independent, their joint probability distribution can be written as the product
  \begin{eqnarray}
    \label{eq.psf}
    p (\boldsymbol{s} \vert f)
    = \prod\nolimits_{j=1}^N p(s_i,f) ,
  \end{eqnarray}
  where we used the notation $\boldsymbol{s}= (s_1,s_2, \dots, s_N)$ and $p(s_i,f)$ is given by Eq.\ (\ref{eq.NNDComp}).  The assumption, actually, restricts the applicability of the approach to short-range correlation functions.
  According to Bayes' theorem the posterior distribution of the
  parameter $f$ for a given sequence of spacings $\boldsymbol{s}$ is
  given as
  \begin{equation}
    P(\boldsymbol{s}\vert f)
    =
    \frac {p(\boldsymbol{s}\vert f)\mu(f)}
    {\mathcal{N}(\boldsymbol{s}) }
\label{PsBayes}
  \end{equation} 
  where $\mu(f)$ is the prior distribution of $f$ and $\mathcal{N}
  (\boldsymbol{s})$ is the normalization constant.  Using Jeffrey's
  rule~\cite{Jeffrey1946, Jeffrey1961, Harney2003}, an approximate expression was derived for $\mu (f)$ in Ref.~\cite{Abul2004}, $\mu(f)=1.975 - 10.07f + 48.96f^2 - 135.6f^3+205.6 f^4 - 158.6f^5 + 48.63f^6$. To determine the chaoticity parameter $f$, we computed the joint
  probability distribution of the spacings $p(\boldsymbol{s}\vert f)$ for $0\leq f\leq 1$ using Eq.~(\ref{eq.NNDComp}). The best-fit value of $f$ was then obtained as the mean value $\overline{f}=\int_0^1fP(\boldsymbol{s}\vert  f){\rm d}f$, which gives the fraction of subspectra that exhibit GOE behavior with variance $\sigma$, $f=\overline{f}\pm\sigma$, where $\sigma^2=\int_0^1(f-\overline{f})^2P (\boldsymbol{s}\vert f){\rm d} f$ provides a measure for the uniformity of chaoticity in the ensemble of subspectra. This yielded the values of $\overline{f}$ and $\sigma$ listed in Table \ref{tab.table1} for the \expt al and calculated spectra. They are in line with those obtained with the RMT model Eq.~(\ref{PGOE}). The corresponding NNSDs are shown as red curves in Fig.~\ref{fig2} and in Fig.~4 of~\refsec{Supplemental}. 

  ~
  {\it Results, discussion, and conclusion}.--
  Table \ref{tab.table1} summarizes the results of the analysis of the recently completed level scheme of \Pb\ at $E_x\ltapprox 6.20$~MeV together with complementary results rom calculations. For the \expt al and the calculated spectra of the nuclear models with M3Y and KB interactions denoted ``all'' in Table \ref{tab.table1}, the chaoticity parameters $f$ and $\lambda$ indicate that the spectral properties are described by GOE, even though there seems to be some small admixture from regular dynamics in the latter two cases. The chaoticity parameters for the SDI, on the contrary, suggest a behavior which is close to Poisson statistics. In a second step the same analysis has been applied to the spectra of energy levels with positive, negative, natural and unnatural parity, respectively. Albeit the sample sizes of $m$ and $N$ are smaller, the chaoticity parameters for the \expt al spectra and for those calculated with the M3Y and KB interactions again are close to the values of GOE statistics. Note, that the parameters are closer to that for a pure GOE for levels with positive and natural parity than for those with negative and unnatural parity. In contrast, all chaoticity parameters obtained for the SDI interaction are compatible with regularity. We may conclude that the SDI, which provides a simple extension of the schematic shell model~\cite{P.mSM_Jolos}, does not describe the underlying interactions in the nucleus \Pb\ correctly. On the other hand, the M3Y and KB interactions are based on realistic, effective nucleon-nucleon interactions. 

However, especially our finding for the levels with unnatural parity differs from those of an analysis of the NNSD based on the same \expt al data~\cite{P.All6.2MeV} in terms of the Brody distribution~\cite{Brody1981} which also depends on a chaoticity parameter~\cite{Munoz2016}. These discrepancies might orginate from the fact that the analysis of Ref.~\cite{Munoz2016} did not include long-range correlations. Nevertheless, we obtain chaoticity parameters close to the values for GOE even when taking into account only the NNSD. To ensure, that this discrepancy does not arise from the unfolding procedure, we furthermore evaluated the ratio distributions using ensembles of subspectra comprising the natural and unnatural parity states, respectively. This analysis corroborated the outcome of the calculations with the RMT model Eq.~(\ref{PGOE}) and the method of Bayesian inference.  

  In conclusion, by analyzing a complete set of levels in \Pb\ with unambiguously assigned spin and parity 
  with the RMT model Eq.~(\ref{PGOE}) and the Bayesian method evidence has been presented for fluctuation properties which
  are consistent with those of random matrices from the GOE, thus for chaoticity of the nuclear system. Similar results where obtained from the analysis of the spectra generated from nuclear model calculations with M3Y and KB interactions. These two models confirm that the chaoticity is caused by a nuclear residual interaction that mixes the \cfg s inextricably in the many-body system. Indeed, e.g., in the M3Y model, the spectral properties of the unperturbed enegery levels, i.e., the diagonal elements of the Hamiltonian $H$, yielded for the chatocity parameters $\bar f=0.19,\ \sigma=0.066$ and $\lambda=0.09,\ \sigma_\lambda=0.010$, respectively, that is they exhibit Poisson statistics, see Fig.~5 of~\refsec{Supplemental}. These values, actually, are close to those for the SDI interaction. Furthermore, the distance between the unperturbed energy levels for each value of $J^\pi$, and the perturbed ones, i.e., the eigenvalues of $H$, is considerably larger than the root-mean square of the off-diagonal interaction matrix elements of $H$; see chapter 4 of Ref.~\cite{Guhr1998}. Thus, the SDI interaction seems to be too weak to induce a sufficient mixture of the individual configurations to yield a chaotic dynamics.
\begin{acknowledgments}
    We thank T. Guhr, T.\ T.\ S.\ Kuo, and H.\ A.\ Weidenm\"uller
    for discussions.  One of us (A.\ R.)\ is grateful to L.\ Mu\~noz
    for providing the article~\cite{Munoz2016} prior to publication.
    This work was supported by the Deutsche Forschungsgesellschaft
    (DFG) within the Collaborative Research Centers 634 and 1245.
    BAB acknowledges U.S. NSF Grant No. PHY-1404442.
\end{acknowledgments}
  \bibliographystyle{unsrt}

\begin{widetext}
\newpage

\section{Supplemental material\label{Supplemental}}
\begin{figure}[h!]
\includegraphics[width=0.4\linewidth]{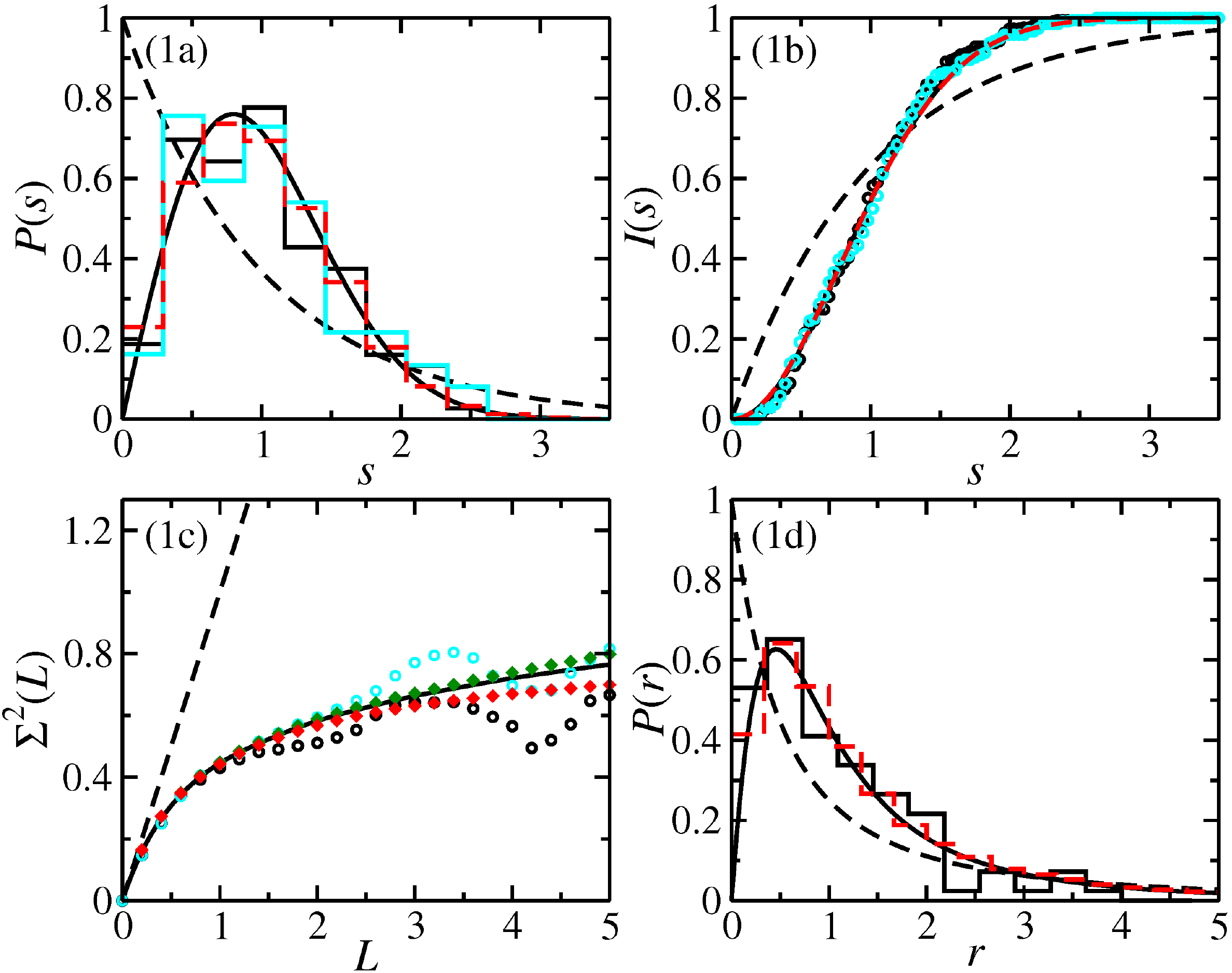}
\includegraphics[width=0.4\linewidth]{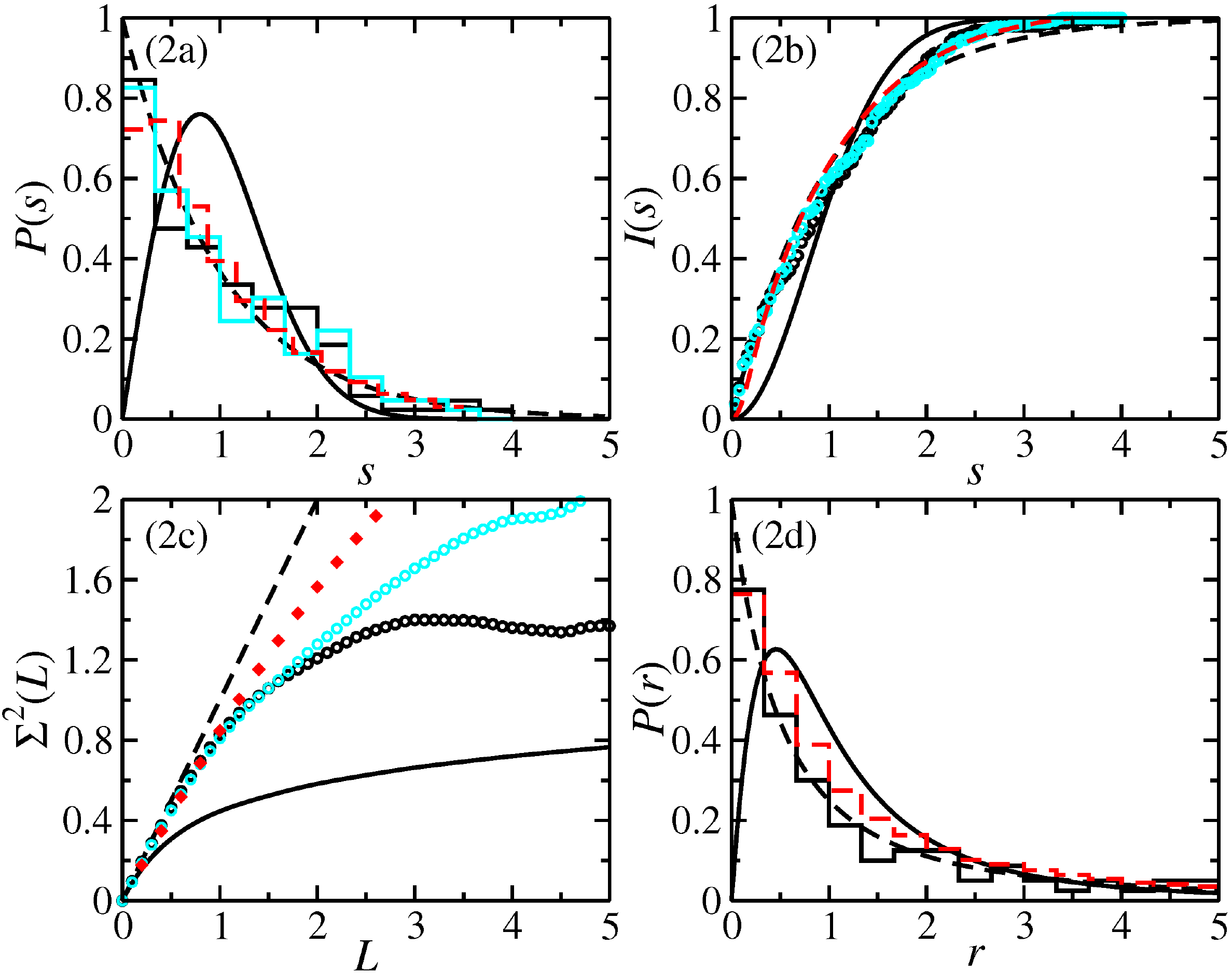}
\includegraphics[width=0.4\linewidth]{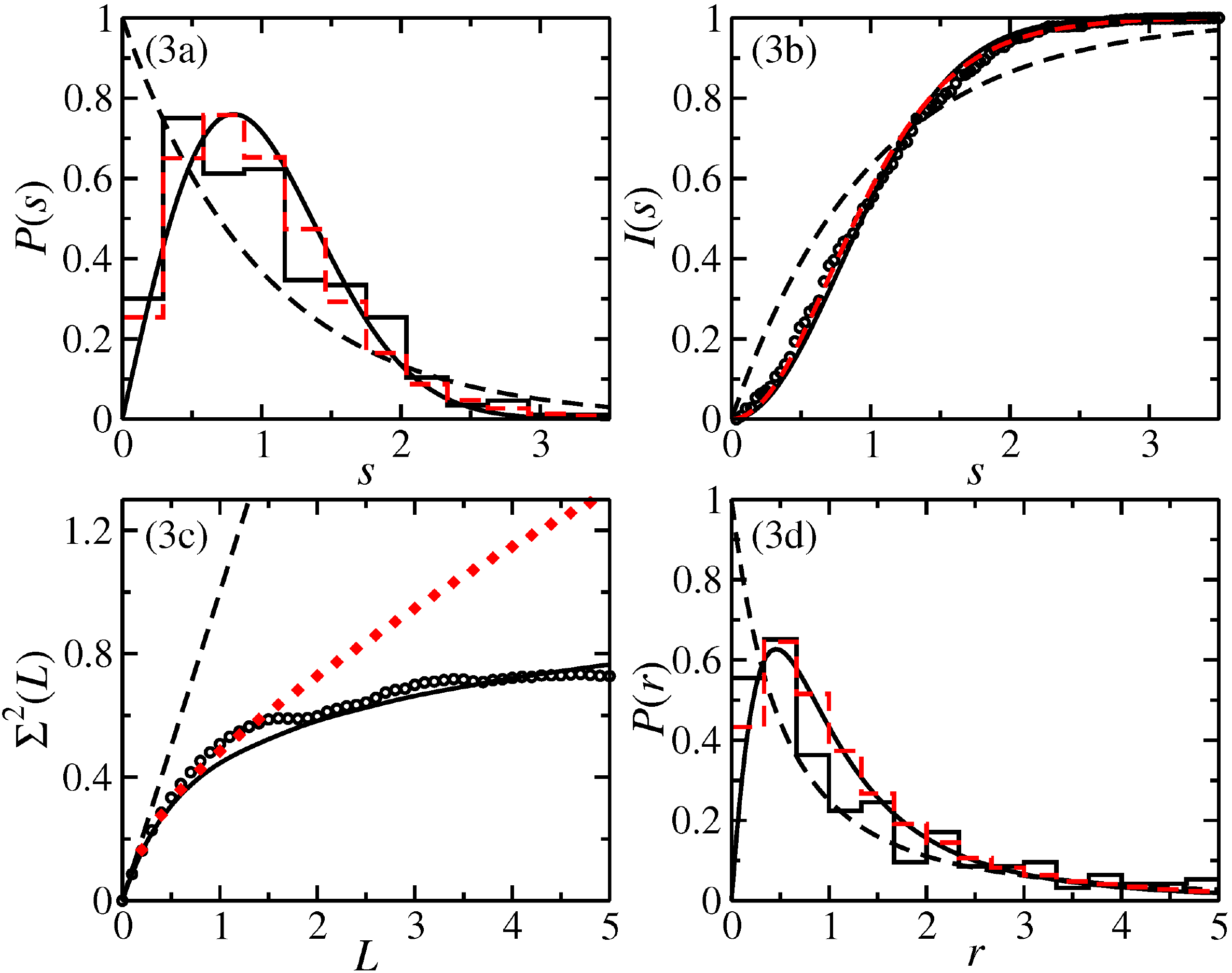}
\includegraphics[width=0.4\linewidth]{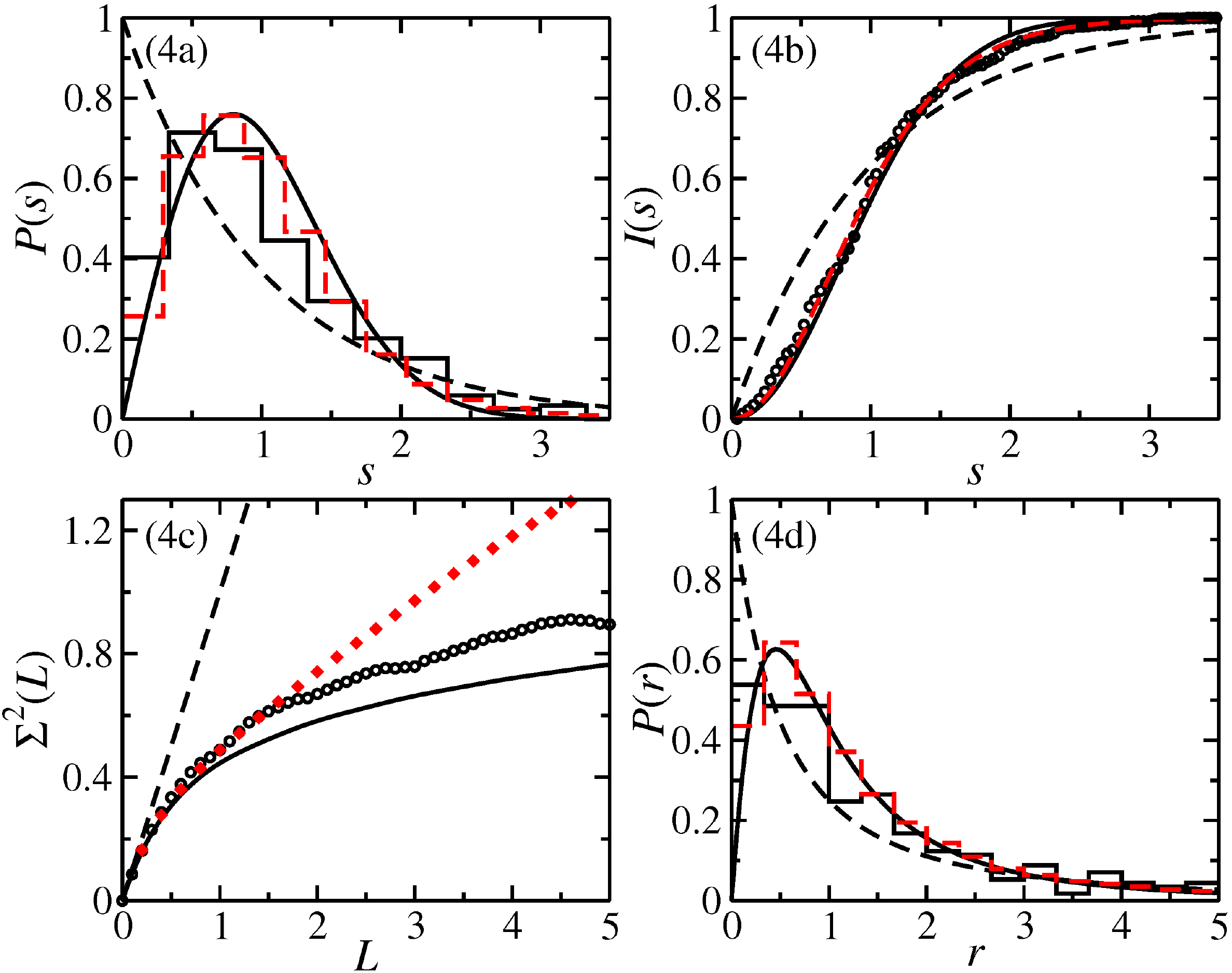}
\caption{(Color online) 
Spectral statistics of \emph{all} experimental energy levels [(1a) - (1d)], and the computed ones using the SDI [(2a) - (2d)], M3Y  [(3a) - (3d)] and KB interactions [(4a) - (4d)]. Shown are the nearest-neighbor spacing distributions $P(s)$, the integrated nearest-neighbor spacing dstributions $I(s)$, the number variance $\Sigma^2(L)$ and the ratio distributions $P(r)$. Black circles and histograms were obtained from the energy levels unfolded with a third-order polynomial. For the experimental data and those computed with the SDI interactions we, in addition, show the statistical measures obtained from the energy levels unfolded with the constant-temperature formula~\cite{Shriner1991} $\bar N (E)=\exp((E-E_0)/T)+N_0$ (cyan circles and histograms). The red diamonds and dashed lines show the corresponding curves for the random matrix model~\cite{Lenz1991,Kota2014} Eq.~(1) interpolating between Poisson ($\lambda=0$) and GOE ($\lambda\gtrsim 1$). The parameter $\lambda$ was determined by fitting the analytical results for the nearest-spacing distribution and the $\Sigma^2$ statistics computed from the random matrix model to the ones obtained for each of the level sequences. The best-fit parameter values and mean-square deviations are given in columns 7 and 8 of Table 1. Dark green diamonds in (1c) show the $\Sigma^2$ statistics obtained from a fit to the one obtained for the experimental data unfolded with the constant-temperature formula. For the remaining statistical measures the curves obtained from both unfolding procedures are barely distinguishable. Therefore we don't show them. Note, that the ratio distributions do not require unfolding and, accordingly, were obtained with the original data.
}
\end{figure}
%
\begin{figure}[h!]
\includegraphics[width=0.4\linewidth]{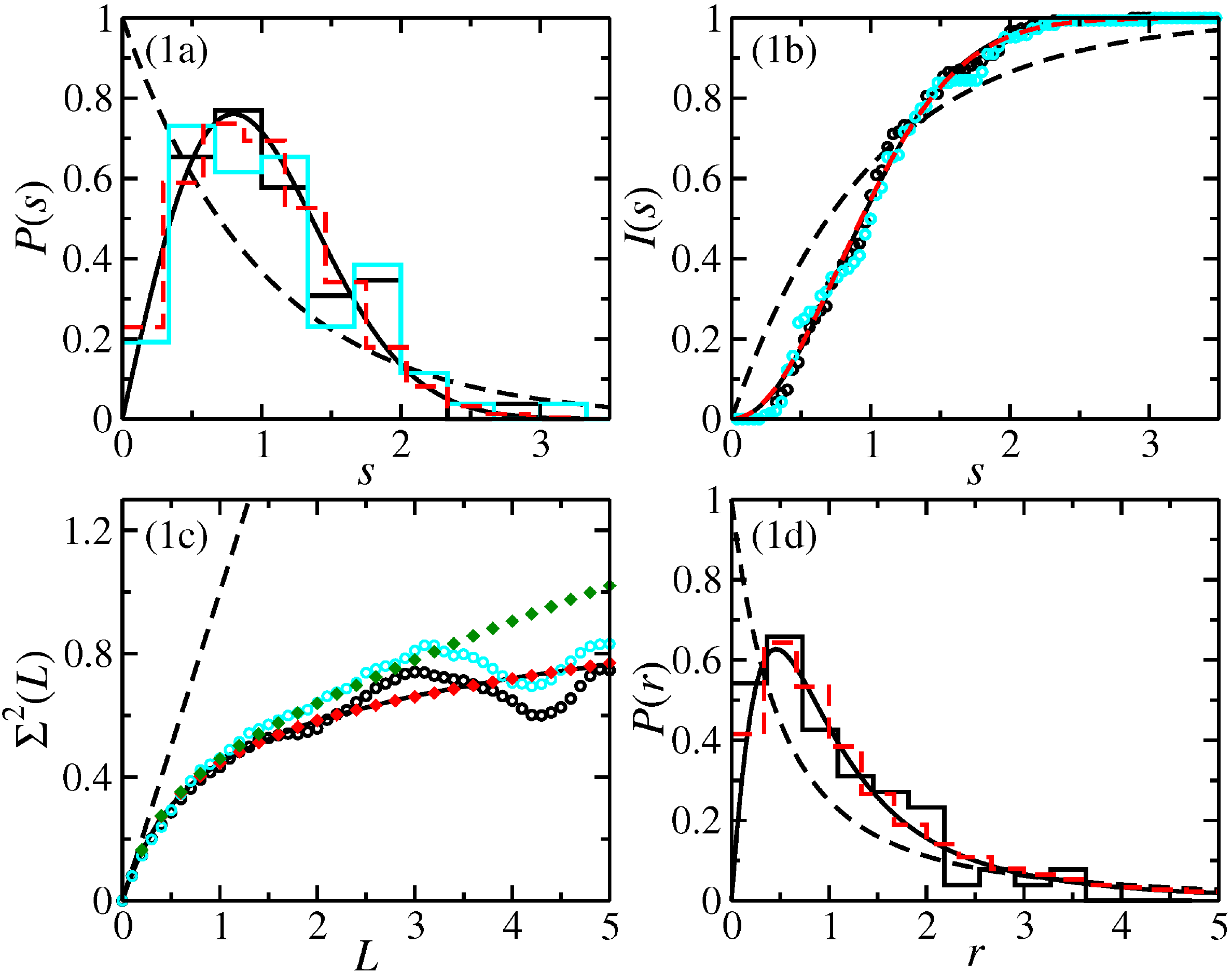}
\includegraphics[width=0.4\linewidth]{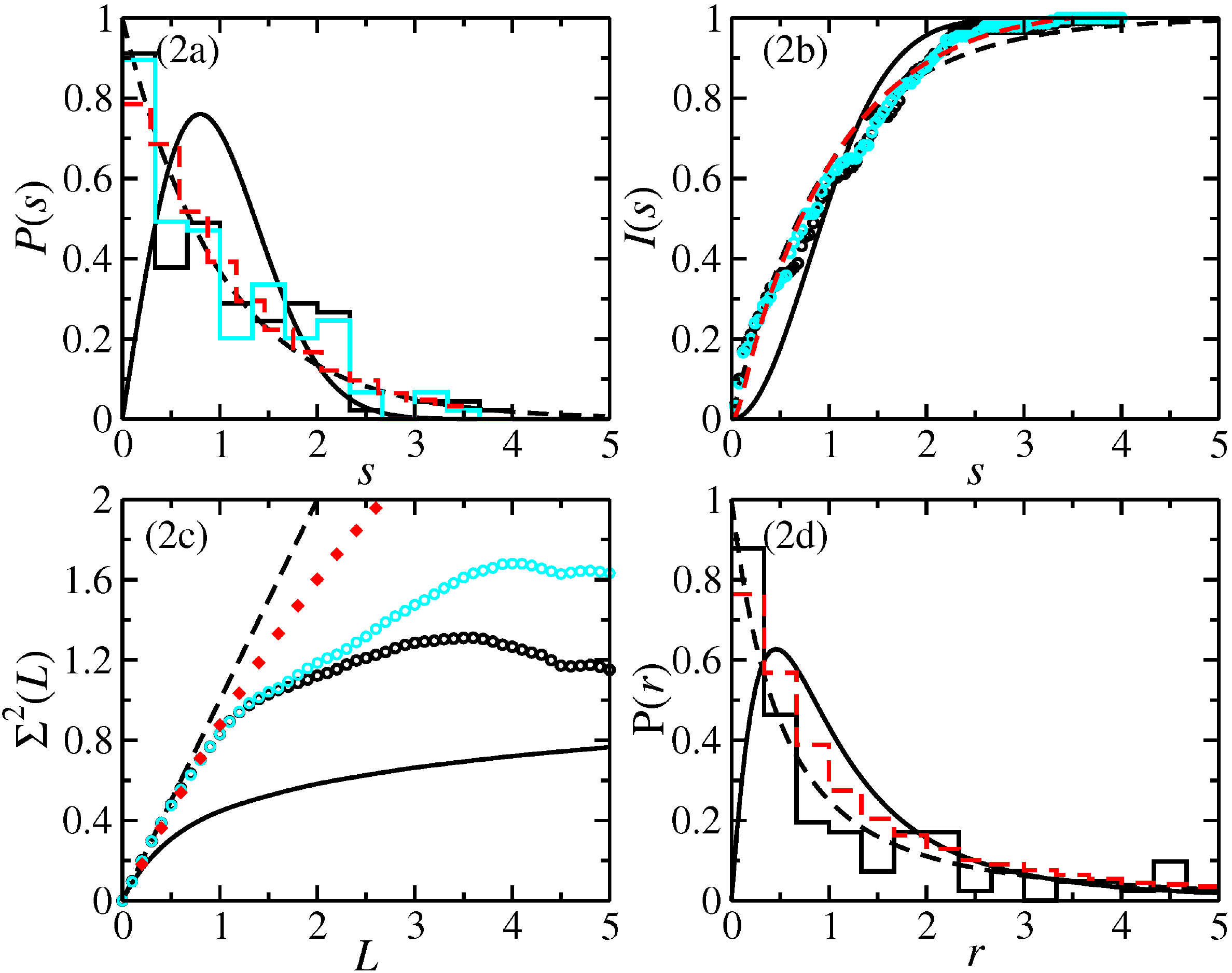}
\includegraphics[width=0.4\linewidth]{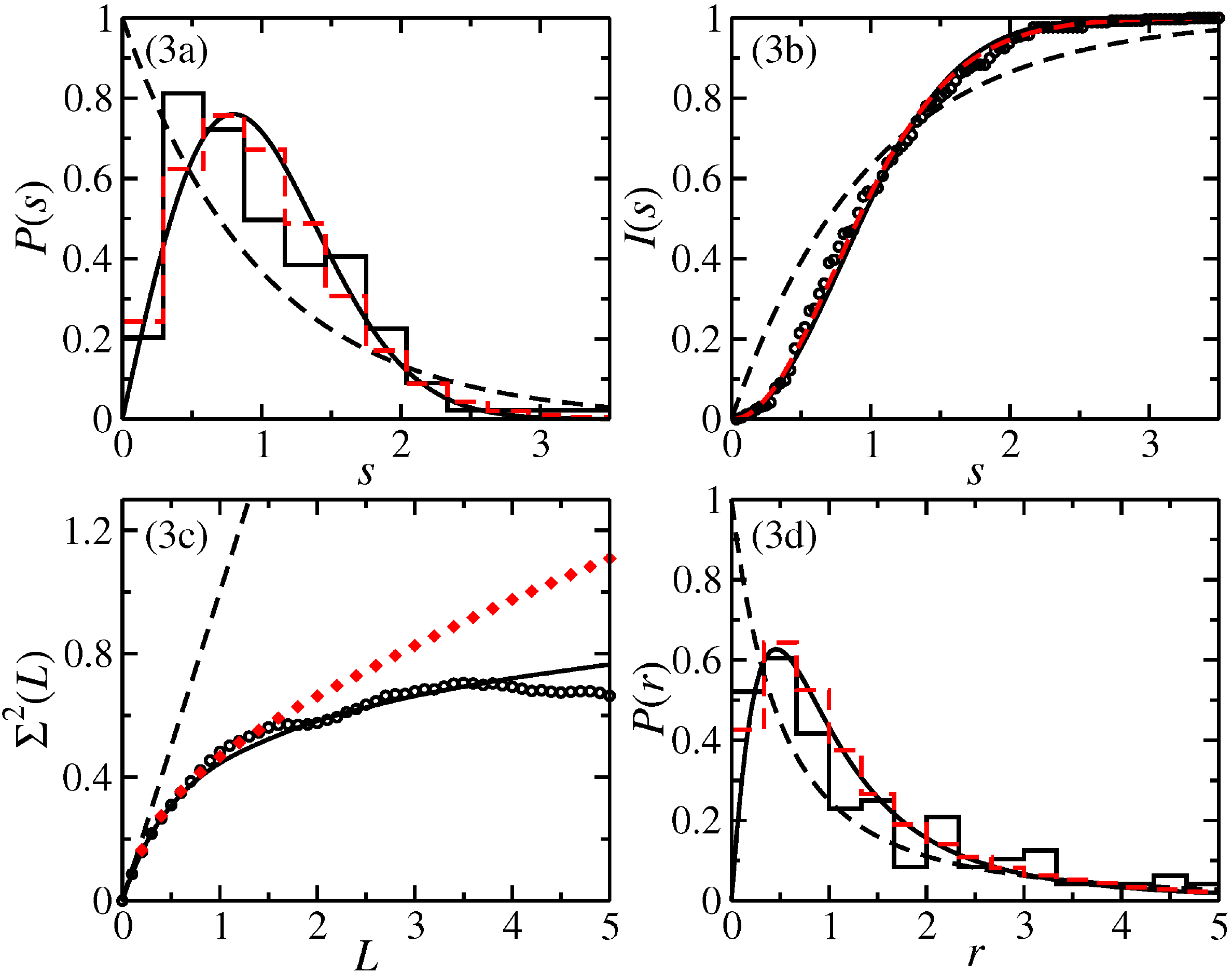}
\includegraphics[width=0.4\linewidth]{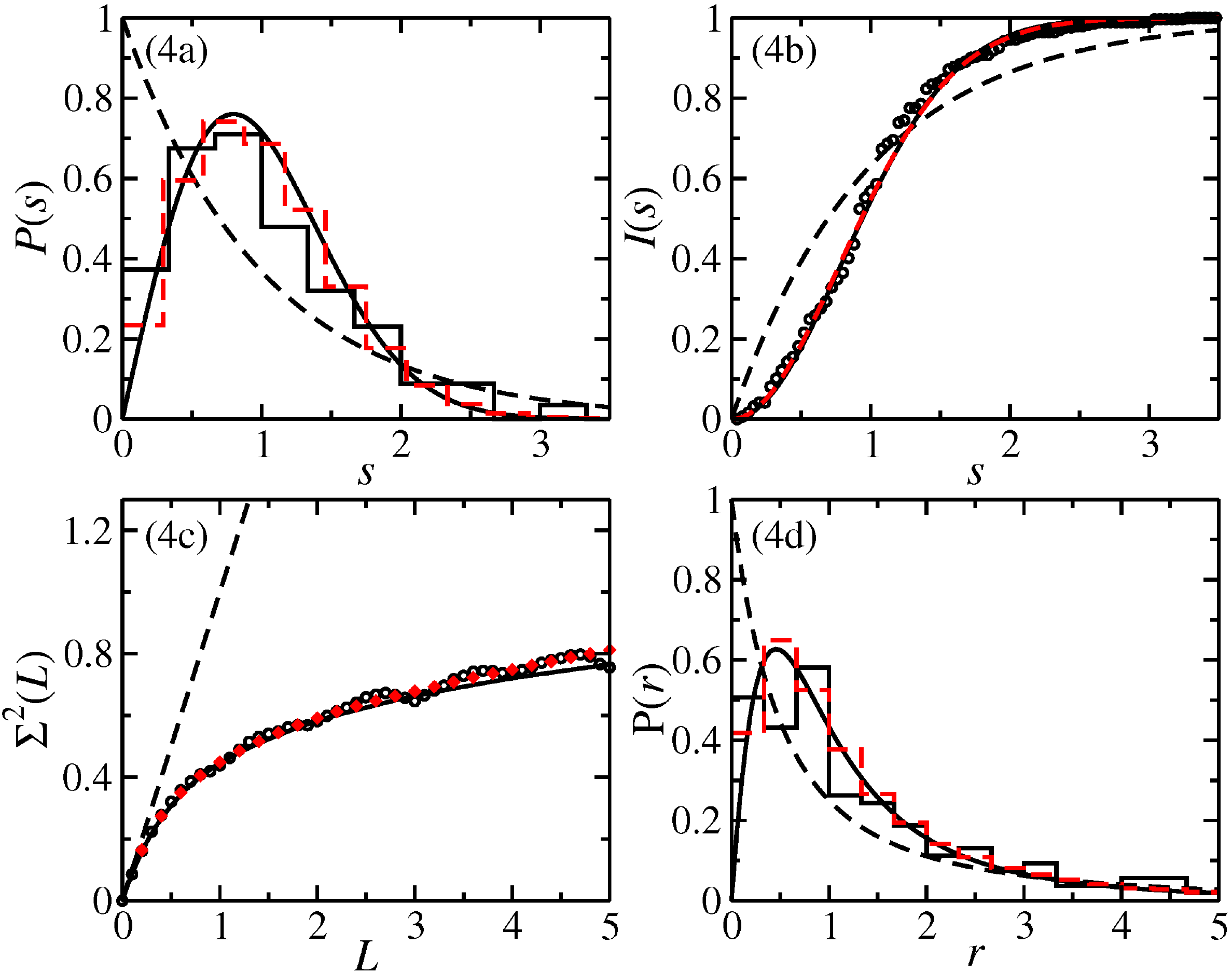}
\caption{(Color online) 
Spectral statistics of the experimental energy levels [(1a) - (1d)], and the computed ones using the SDI [(2a) - (2d)], M3Y  [(3a) - (3d)] and KB interactions [(4a) - (4d)] with \emph{natural} parity. Shown are the nearest-neighbor spacing distributions $P(s)$, the integrated nearest-neighbor spacing dstributions $I(s)$, the number variance $\Sigma^2(L)$ and the ratio distributions $P(r)$. Black circles and histograms were obtained from the energy levels unfolded with a third-order polynomial. For the experimental data and those computed with the SDI interactions we, in addition, show the statistical measures obtained from the energy levels unfolded with the constant-temperature formula~\cite{Shriner1991} $\bar N (E)=\exp((E-E_0)/T)+N_0$ (cyan circles and histograms). The red diamonds and dashed lines show the corresponding curves for the random matrix model~\cite{Lenz1991,Kota2014} Eq.~(1) interpolating between Poisson ($\lambda=0$) and GOE ($\lambda\gtrsim 1$). The parameter $\lambda$ was determined by fitting the analytical results for the nearest-spacing distribution and the $\Sigma^2$ statistics computed from the random matrix model to the ones obtained for each of the level sequences. The best-fit parameter values and mean-square deviations are given in columns 7 and 8 of Table 1. Dark green diamonds in (1c) show the $\Sigma^2$ statistics obtained from a fit to the one obtained for the experimental data unfolded with the constant-temperature formula. For the remaining statistical measures the curves obtained from both unfolding procedures are barely distinguishable. Therefore we don't show them. Note, that the ratio distributions do not require unfolding and, accordingly, were obtained with the original data.}
\end{figure}
%
\begin{figure}[h!]
\includegraphics[width=0.4\linewidth]{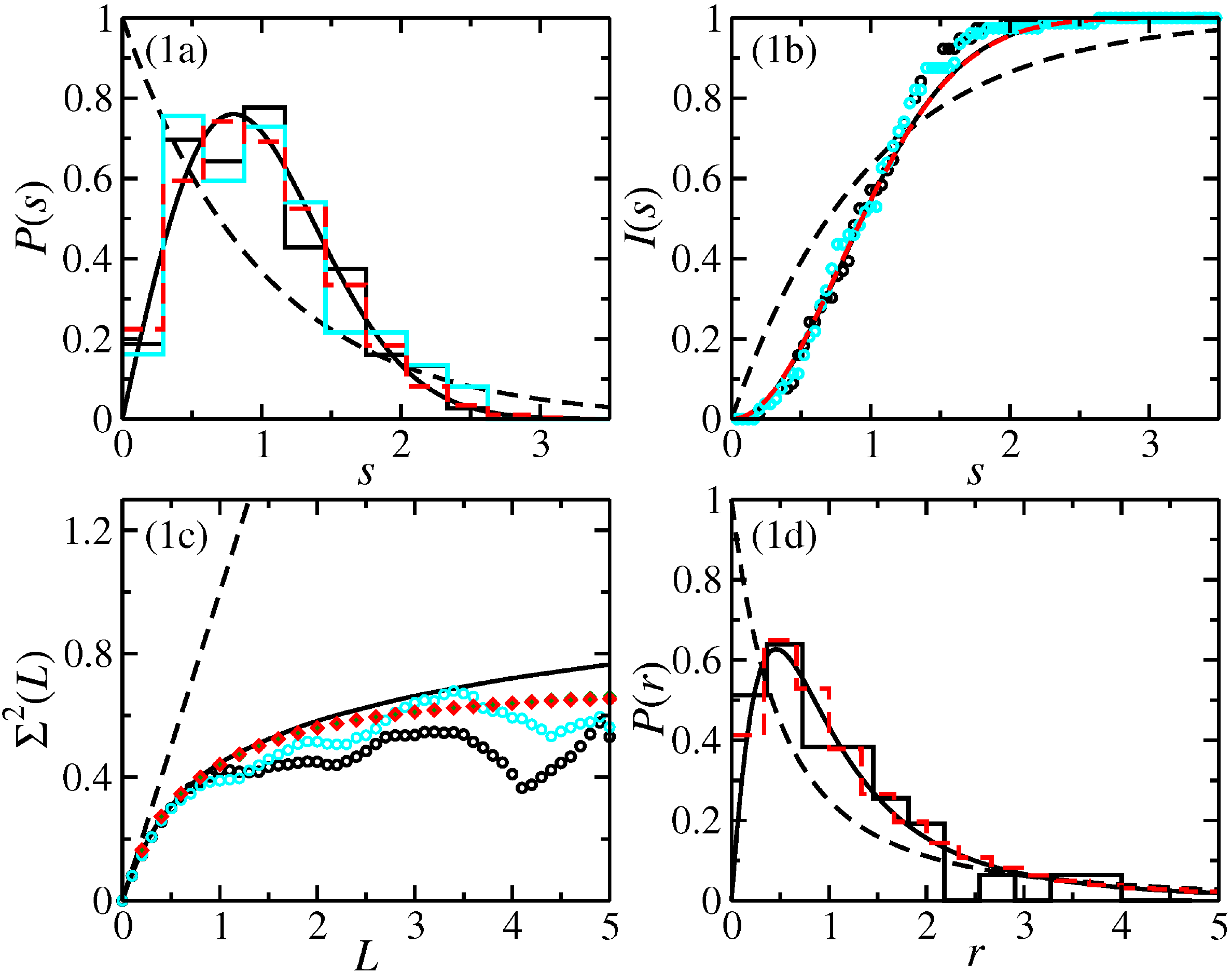}
\includegraphics[width=0.4\linewidth]{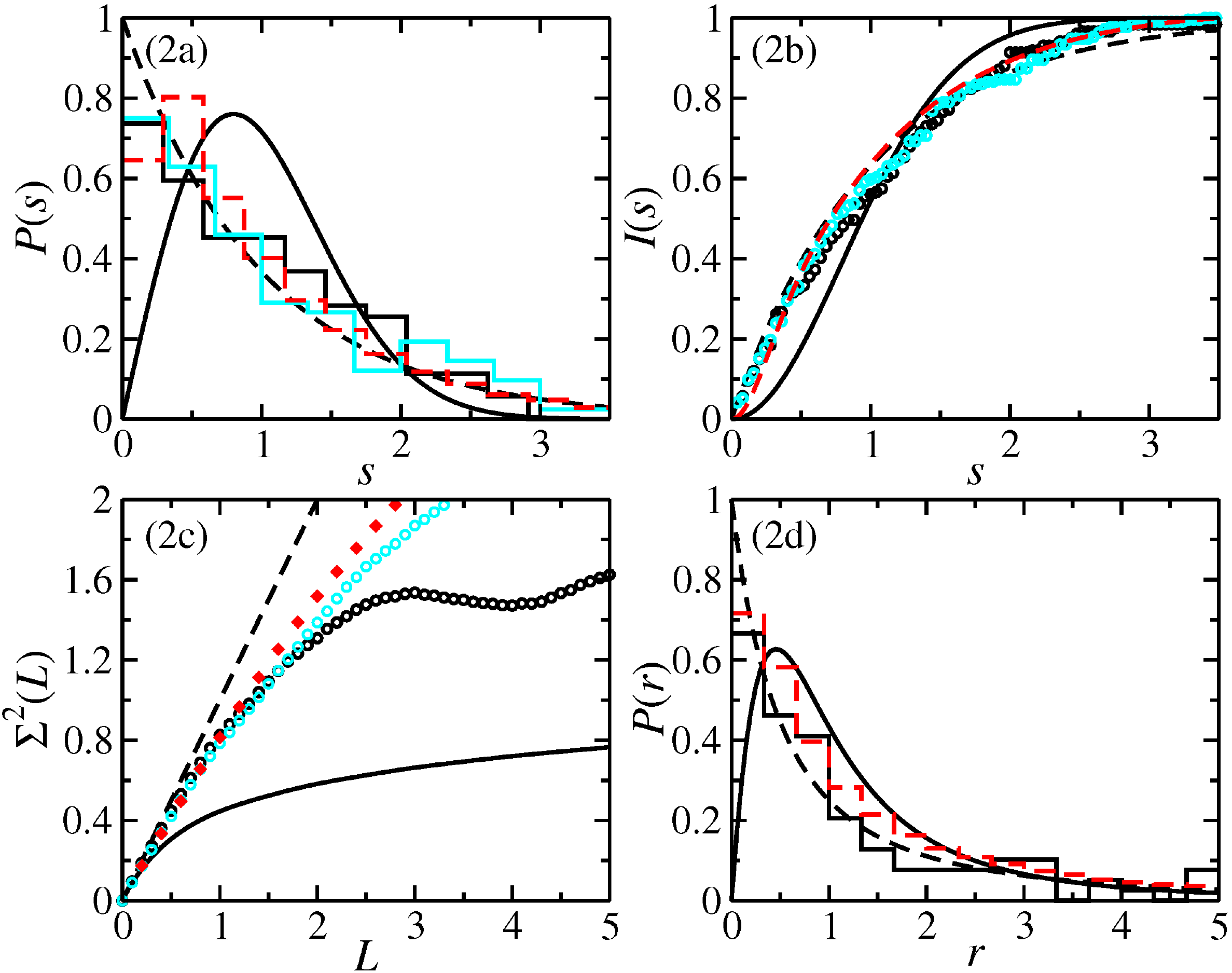}
\includegraphics[width=0.4\linewidth]{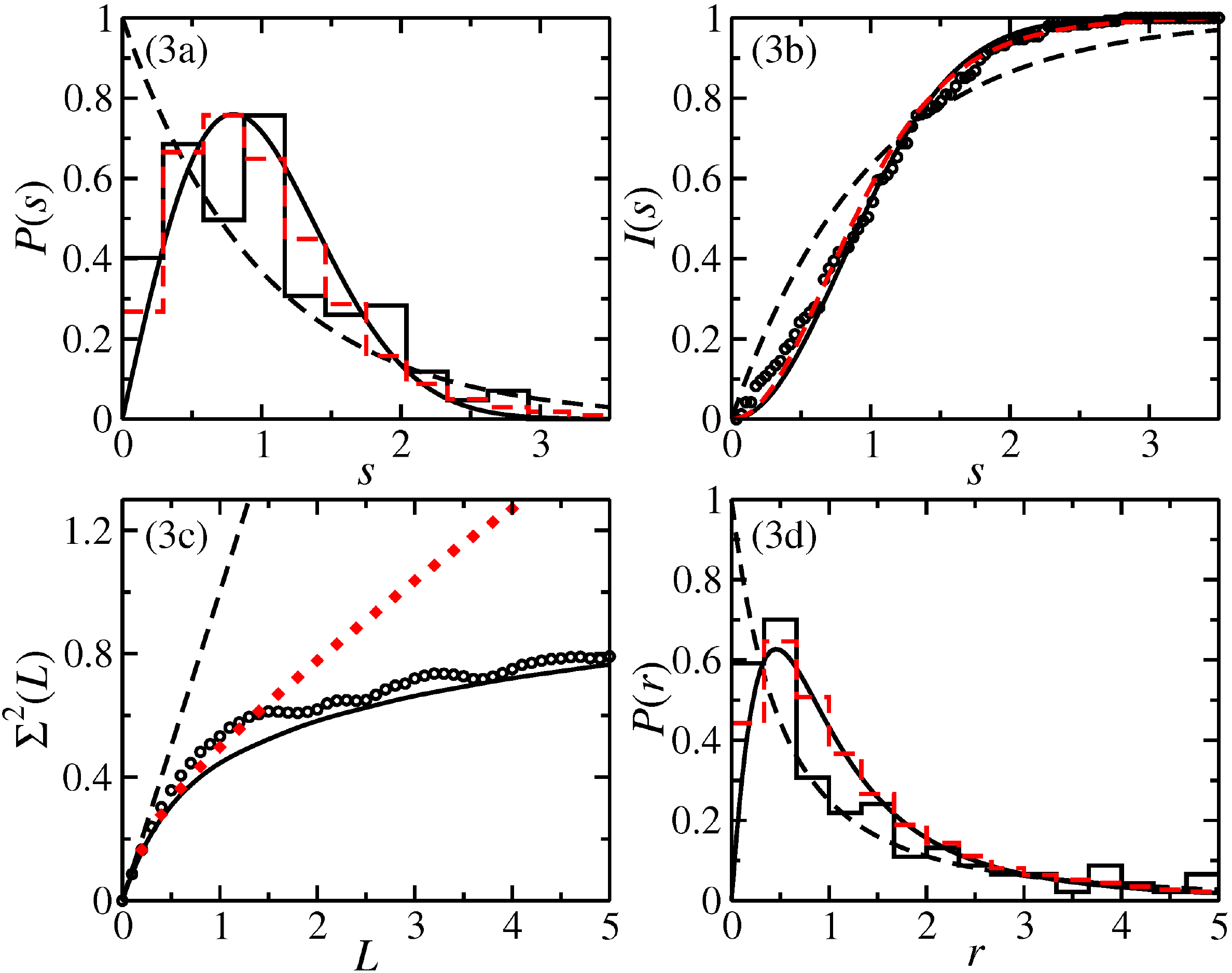}
\includegraphics[width=0.4\linewidth]{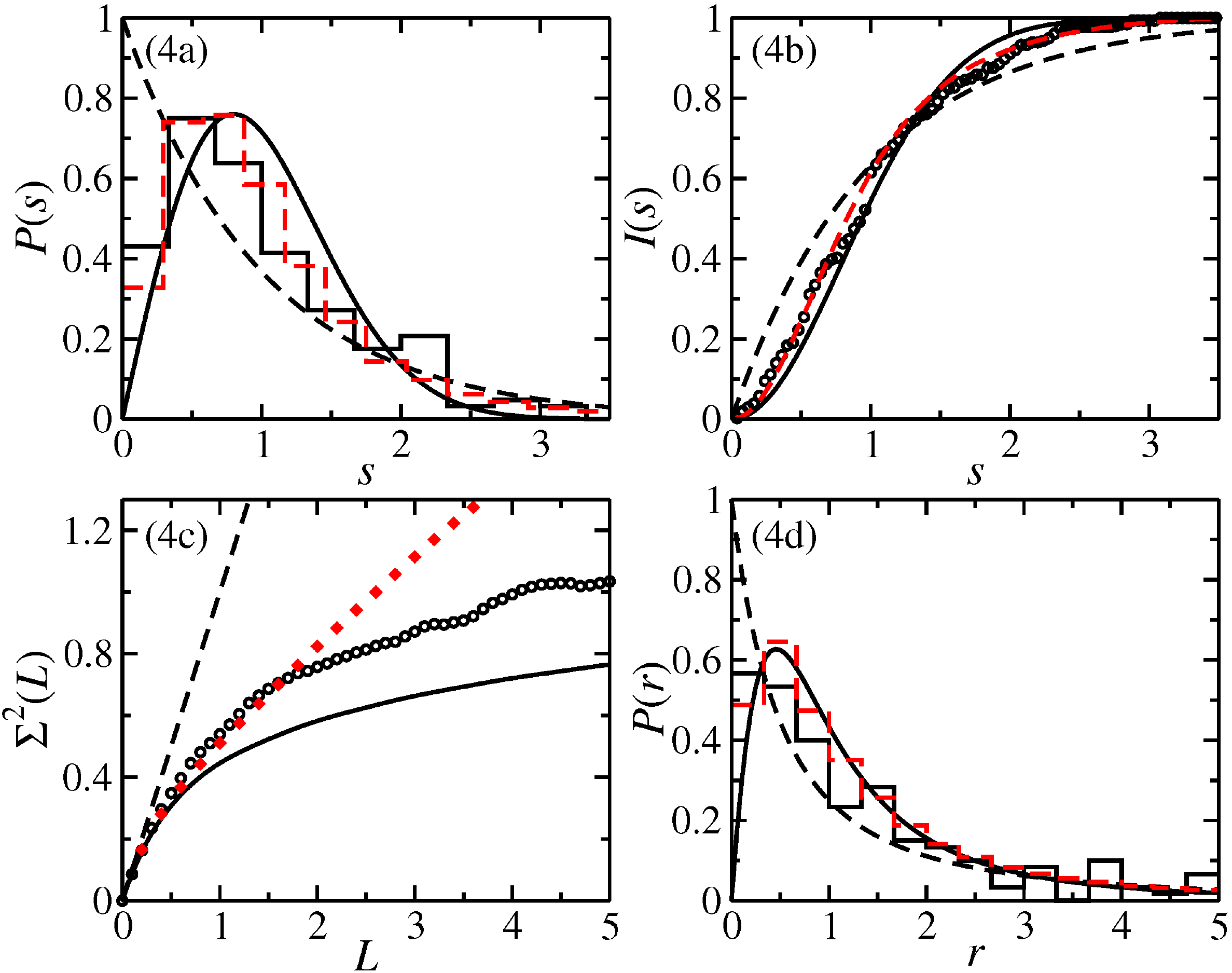}
\caption{(Color online) 
Spectral statistics of experimental energy levels [(1a) - (1d)], and the computed ones using the SDI [(2a) - (2d)], M3Y  [(3a) - (3d)] and KB interactions [(4a) - (4d)] with \emph{unnatural} parity. Shown are the nearest-neighbor spacing distributions $P(s)$, the integrated nearest-neighbor spacing dstributions $I(s)$, the number variance $\Sigma^2(L)$ and the ratio distributions $P(r)$. Black circles and histograms were obtained from the energy levels unfolded with a third-order polynomial. For the experimental data and those computed with the SDI interactions we, in addition, show the statistical measures obtained from the energy levels unfolded with the constant-temperature formula~\cite{Shriner1991} $\bar N (E)=\exp((E-E_0)/T)+N_0$ (cyan circles and histograms). The red diamonds and dashed lines show the corresponding curves for the random matrix model~\cite{Lenz1991,Kota2014} Eq.~(1) interpolating between Poisson ($\lambda=0$) and GOE ($\lambda\gtrsim 1$). The parameter $\lambda$ was determined by fitting the analytical results for the nearest-spacing distribution and the $\Sigma^2$ statistics computed from the random matrix model to the ones obtained for each of the level sequences. The best-fit parameter values and mean-square deviations are given in columns 7 and 8 of Table 1. Dark green diamonds in (1c) show the $\Sigma^2$ statistics obtained from a fit to the one obtained for the experimental data unfolded with the constant-temperature formula. They lie on top of the red diaminds. For the remaining statistical measures the curves obtained from both unfolding procedures are barely distinguishable. Therefore we don't show them. Note, that the ratio distributions do not require unfolding and, accordingly, were obtained with the original data.}
\end{figure}
%
\begin{figure}[h!]
\includegraphics[width=0.7\linewidth]{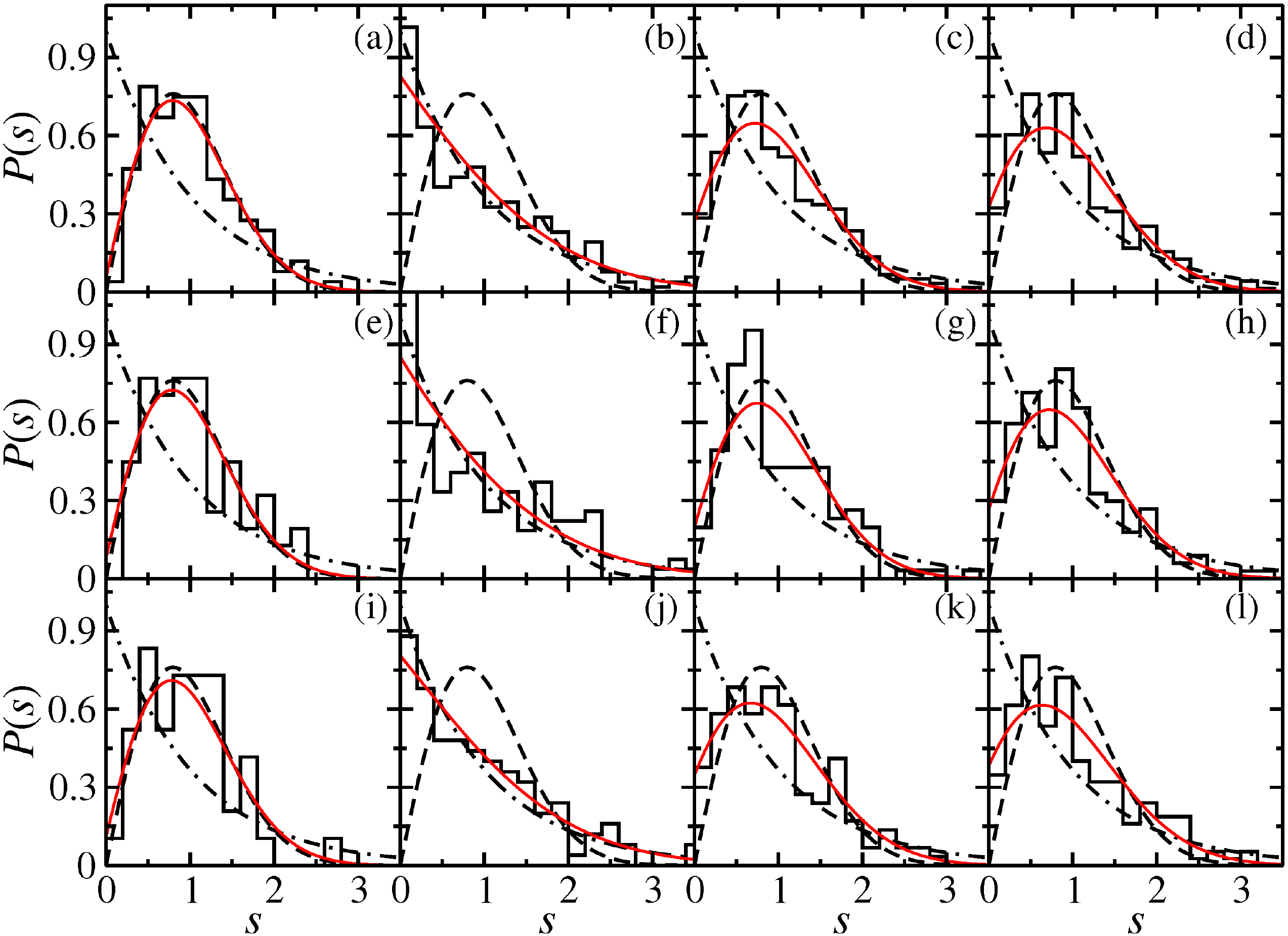}
\caption{(Color online)
Nearest-neighbor spacing distributions of the experimental level spacings (first column) and of the computed ones using the SDI (second column) , M3Y (third column), and KB (fourth column) interactions (histograms). They are compared to the Poisson (dash-dotted line) and the GOE (dashed line) distribution. The full curves in red were determined with the method of Bayesian inference [Eqs.~(3),~(5) and~(6)]. The corresponding values of the chaoticity parameter and the variances are given in the fifth and sixth column of Table 1, respectively. The histograms were obtained by taking into account all levels [(a)-(d)], levels with natural parity [(e)-(h)] and levels with unnatural parity [(i)-(l)]. 
}
\end{figure}
%
%
\begin{figure}[h!]
\includegraphics[width=0.5\linewidth]{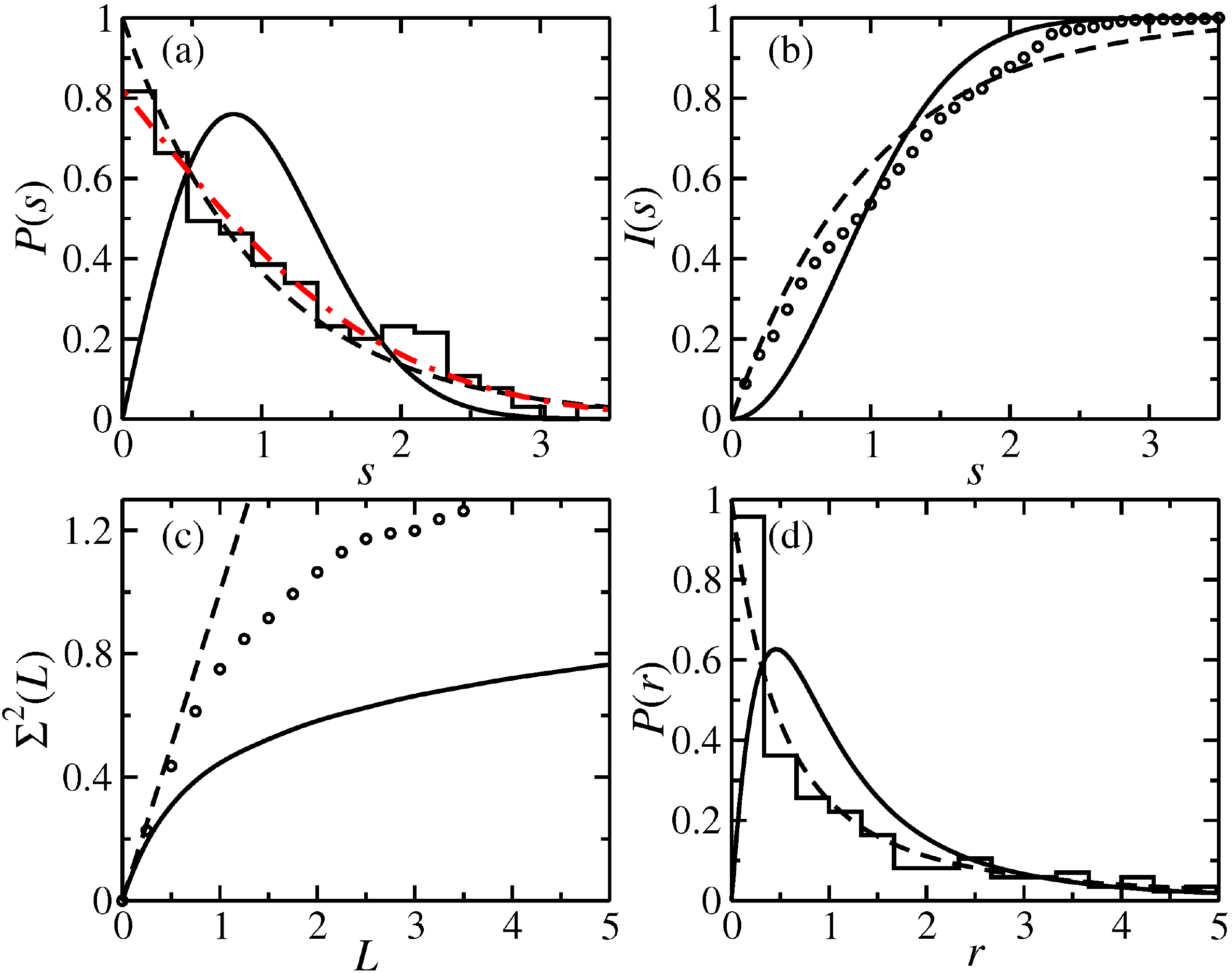}
\caption{(Color online)
Same as shown in Fig. 3, [(3a)-(3d)]. However, only the unperturbed part of the Hamiltonian associated with the M3Y model was taken into account. The spectral properties are well described by Poissonian level statistics. The red dash-dotted curve was determined with the method of Bayesian inference.
}
\end{figure}
%
%
\begin{figure}[h!]
\includegraphics[width=0.5\linewidth]{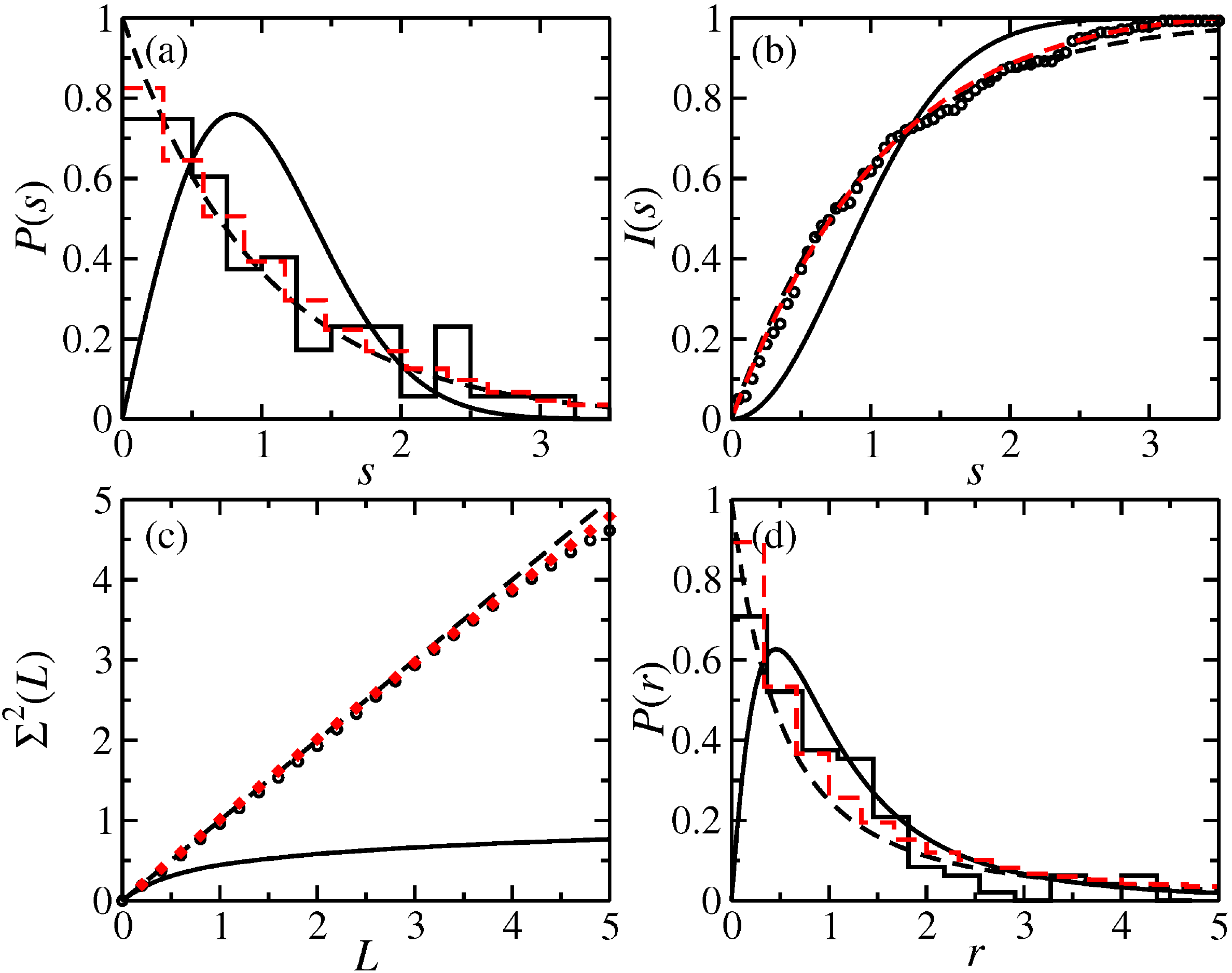}
\caption{(Color online)
Spectral properties for a sequence of all experimental energy levels constructed by combining them irrespective of their spin and parity. As expected, they are well described by Poissonian level statistics.
}
\end{figure}
\end{widetext}
\end{document}